# THE ORIGINS AND THE EARLY EVOLUTION OF QUASARS AND SUPERMASSIVE BLACK HOLES[*]

S. G. DJORGOVSKI

*California Institute of Technology*
*Pasadena, CA 91125, USA*

MARTA VOLONTERI

*Dept. of Astronomy, University of Michigan*
*Ann Arbor, MI 48109, USA*

VOLKER SPRINGEL

*Max Planck Institute for Astrophysics*
*Garching, 85741, Germany*

VOLKER BROMM

*Dept. of Astronomy, University of Texas*
*Austin, TX 78712, USA*

GEORGES MEYLAN

*Laboratory of Astrophysics, Ecole Polytechnique Fédérale de Lausanne (EPFL),*
*Observatoire de Sauverny, CH-1290 Versoix, Switzerland*

The relationship between galaxies and supermassive black holes (SMBH) found in their cores plays a key role in the formation and evolution of both of these major constituents of the universe, as well as the evolution of the intergalactic medium. Neither can be fully understood on their own, and studies of galaxy and SMBH co-formation and co-evolution are now among the central topics of research in cosmology. Yet the very origins, and the early growth phases of the SMBH are still not firmly established. We review our current understanding of the relevant processes and their astrophysical and cosmological context, with an emphasis on the observability of the SMBH growth mechanisms at high redshifts, and their leftover progeny at low redshifts.

## 1   Introduction

Who knows what darkness lurks in the hearts of galaxies? Astronomers know. But how did it get there? That is a bit more mysterious.

We now believe that compact, massive, dark, and probably relativistic objects – commonly known as black holes (BH) – exist in the nuclei of many, if not all large galaxies today (the extended family of dwarf galaxies seems to be exempt, but not excluded from this demographic propensity). The "compact, massive, dark, and probably relativistic" attributes suffice to account for all of the observed astrophysical

---

phenomenology associated with these objects, regardless of the precise nature of their physics, as one approaches their effective boundaries with the classically implied infinite gravitational potentials, and probably some profound quantum gravity physics yet to come. The subject has a rich and growing literature, including many excellent reviews elsewhere in this volume, to which we refer for further information and references. In any case, we can discuss observed manifestations of BHs at larger scales and within their host environments even if we do not yet really understand fully what happens near, or within the radii corresponding to their classical event horizons.

In keeping with a common practice, we will refer to BHs with masses $M_{BH} \gtrsim 10^5 \, M_\odot$ as "supermassive" (SMBH), those with masses $M_{BH} \sim 10^2 - 10^5 \, M_\odot$ as "intermediate" (IMBH), distinguishing them from the "stellar" mass range, $M_{BH} \sim 10^0 - 10^2 \, M_\odot$ ; these boundaries are of course somewhat arbitrary, but they do at least separate the galaxian from the stellar mass regimes.

Accretion of material to SMBHs, with a corresponding release of binding energy comparable to a substantial fraction of the fuel's rest mass energy, has been pinpointed as the most likely mechanism to power quasars (QSOs) and other active galactic nuclei (AGN) since 1960's (Salpeter 1964, Zel'dovich 1964). Thus, observing AGN means observing the growth of their SMBH, at least through the accretion mechanism.

For a typical QSO luminosity of $L_Q \sim 10^{12} \, L_\odot$, sustained over an estimated typical period of activity of $t_Q \sim 10^7 - 10^8$ yr, the total energy release is $E_Q \sim 10^{60} - 10^{61}$ erg, which is up to two orders of magnitude higher than the binding energy of a typical galaxy. Thus, even if a small fraction of this energy is coupled to the ISM of the host, it could exert a substantial mechanical feedback. A similar argument can be made about the ionizing luminosity of the QSO, and its effect on the ambient star formation, with the effective Stromgren sphere radii of the orders of Mpc or larger. Thus, AGN feedback is now considered as an essential component of the galaxy formation and evolution process, as we discuss in more detail below.

This was demonstrated powerfully by the existence of excellent correlations between SMBH masses and their host galaxy properties (luminosities, stellar masses, velocity dispersions, dark halo masses) which are measured on radial scales $\sim 10^8 - 10^9$ times the Schwarzschild radii (for reviews and references, see, e.g., Kormendy & Richstone 1995, Ferrarese 2004, Ferrarese & Ford 2005, etc.). The exact mechanisms leading to these remarkable correlations are not yet firmly established, although dissipative merging and infall, probably regulated by the AGN feedback, must play a role.

While SMBH-powered AGN are fascinating objects in their own right, and have been long used as cosmic laboratories of relativistic astrophysics, they can also serve as valuable probes of the processes of early galaxy and structure formation, as well as their interplay with the IGM and early stages of the "cosmic web". This is especially true for the highest redshift quasars, and their progenitors. The exact physical mechanisms of the SMBH origins and early growth, presumably occurring in the earliest stages of galaxy formation during the reionization era, and their relative importance in the overall evolutionary picture of the QSO population, are one of the more vibrant areas of research in cosmology today.

In this review we address some of these issues, but we note that the subject is changing rapidly, as it befits any frontier area in science. Numerous good reviews of these and related topics include, e.g., Haiman & Quataert (2004), Madau (2005, 2007), Rees & Volonteri (2007), Dokuchaev et al. (2007), Volonteri (2008), and many more can be found, e.g., in the conference proceedings edited by Ho (2004), Barger (2004), Merloni (2005), Colpi et al. (2006), Karas & Matt (2007), Livio & Koekemoer (2008), etc.

## 2   Quasars at Large Redshifts

Quasars remain one of the most powerful probes of the early universe, providing measurements and constraints for reionization, early structure formation, and early chemical enrichment. For reviews and references, see, e.g., Fan (2006), Djorgovski (2005) or Djorgovski et al. (2006).

Absorption spectra of QSOs at z > 5.5 provide some of the key measurements for the evolution of the IGM transmission at the end of the cosmic reionization; see, e.g., Fan et al. (2006) for a review and references. This includes direct measurements of the optical depth of IGM absorption, and estimates of the radii of QSO Stromgren spheres, which constrain the neutral hydrogen fraction in the regimes $x_{HI} \sim 10^{-4} - 10^{-2}$, and $\sim 10^{-1}$, respectively. Metallic line absorbers seen in their spectra constrain the early chemical evolution of the IGM.

QSOs out to z ~ 6 seem to be powered by SMBHs with masses reaching a few × $10^9$ $M_{\odot}$ (Vestergaard 2004, McLure & Dunlop 2004), implying an early and extremely efficient mass assembly; see also Kurk et al. (2007), and Jiang et al. (2007). This is not an easy thing to arrange, as we discuss in more details below.

Abundance analysis of QSO spectra reaching out to z ~ 6 implies a high degree of chemical evolution on their hosts, with metallicities up to ~ 10 $Z_{\odot}$ and Fe/Mg ratios indicative of a very early chemical enrichment by type Ia SNe (Dietrich et al. 2003ab). Intriguingly, large Fe/Mg ratios are also predicted for the element production by the Pop. III pair-instability SNe (see, e.g., Bromm et al. 2003).

This strong and rapid chemical enrichment is consistent with the growing number of detections of thermal dust continuum and molecular lines from high-*z* QSOs, indicative of a vigorous star formation in these objects, or their immediate vicinity (Omont et al 2001, 2002; Carilli et al. 2001; Bertoldi et al. 2003; Petric et al. 2003; Wang et al. 2007, Maiolino et al. 2007, etc.). The bolometric luminosities reach $L_{IR} \sim 10^{13}$ $L_{\odot}$, and the inferred molecular gas masses ~ $10^{10}$ $M_{\odot}$ (and these are *unobscured* QSOs – otherwise they would not have been detected in the optical = restframe UV surveys).

All of these observations represent non-trivial constraints for theoretical models, which become sharper as the QSO redshift increases. The current QSO record redshift holder is SDSS 1148+5251 at *z* = 6.41 (Fan et al. 2003), which is in many ways typical:

It is powered by a ~ $3 \times 10^9$ $M_\odot$ SMBH (Willott et al. 2003), has a super-Solar metallicity and Fe/Mg ratios suggestive of a starburst activity reaching out to z ~ 10 (Barth et al. 2003), and has CO detections indicative of a massive star formation rate (Bertoldi et al. 2003, Walter et al. 2003). In other words, it is clearly an already well-evolved object.

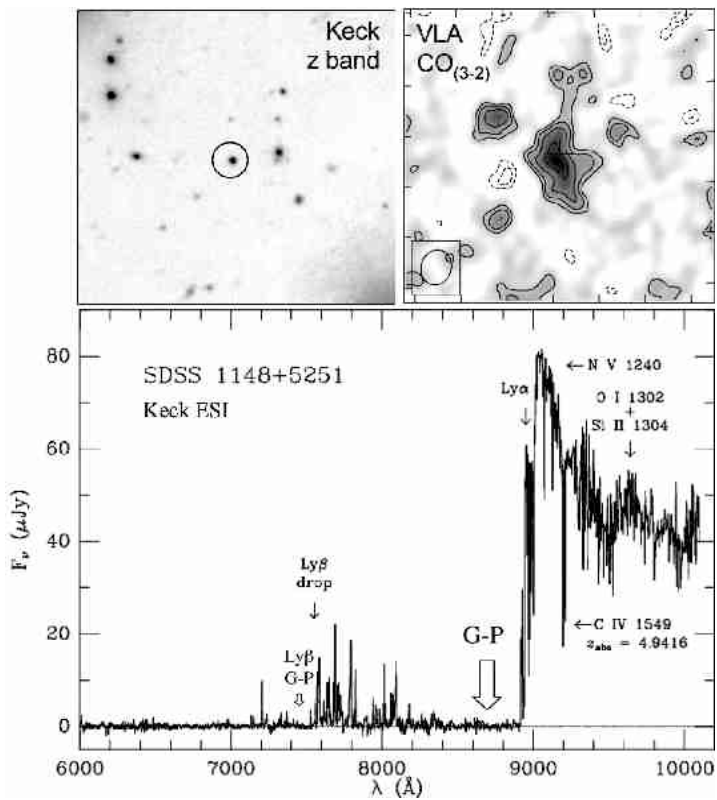

**Figure 1.** The most distant QSO currently known, SDSS 1148+5251, at z = 6.42 (Fan et al. 2003). The top left panel shows an image obtained at the Keck 10-m telescope in the Gunn z band ($\lambda_{obs}$ ~ 900 nm, $\lambda_{rest}$ ~ 120 nm); the QSO is circled. The top right panel shows the VLA radio image in the CO(3-2) line, from Walter et al. (2004); this is indicative of a strong star forming activity in the host galaxy. The bottom panel shows a spectrum obtained at the Keck 10-m telescope, with a very prominent Gunn-Peterson trough just blueward of the Lyα line, illustrating the utility of high-redshift QSOs as probes of the end of the reionization era.

Regardless of the actual physical mechanism(s) by which these SMBHs are assembled, and which we discuss below, at least ~ $10^8$ $M_\odot c^2 \approx 10^{62}$ erg worth of energy must be radiated over a period of ~ 0.5 − 1 Gyr. Thus, progenitors of QSOs we already observe out to z ~ 6.4 must exist at even higher redshifts, and must be very luminous objects themselves, with $\langle L_{protoQ} \rangle > 10^{12} \, L_\odot$ in that time interval. This is also in agreement with reasonable extrapolation of the measured, evolving bright QSO luminosity function (QLF) out to such redshifts; see, e.g., Fan et al. (2001).

The problem in pushing to higher redshifts is that the Ly$\alpha$ line and the associated flux drop used to color-select QSOs move outside the CCD wavelength range for $z > 6.5$ or so. Since the most luminous QSOs are extremely rare, especially at such high $z$'s, a large, moderately deep, panoramic IR sky survey is needed. The UKIDSS survey (Warren et al. 2007, Lawrence et al. 2007; see http://www.ukidss.org/), combined with the modern panoramic optical sky surveys such as SDSS or PQ (Djorgovski et al. 2008; see http://palquest.org), offers an opportunity to discover bright QSOs at $z > 6.5$. For some early attempts, see Venemans et al. (2007), or Glikman et al. (2008).

It is also possible that optical/IR identifications of radio or X-ray sources may lead to discoveries of QSOs at $z > 6.5$, especially with the advent of the next generation of radio observatories, such as ALMA, EVLA, ATA, or SKA (see, e.g., Haiman et al. 2004).

## 3   Possible Formation Mechanisms for High-Redshift SMBHs

Where did these $10^9$ M$_\odot$ SMBHs powering high-$z$ QSOs come from? In a now standard "concordance cosmology", there is $\sim 0.7 - 0.8$ Gyr available from the highest redshifts where the formation of BH seeds may start ($z \sim 30$, say), and the epoch at which we observe these luminous QSOs at $z \sim 6$. If we start with a seed BH with $M_{BH} \sim 10^2$ M$_\odot$, we need $\sim 16$ $e$-folding times to grow a SMBH with $M_{BH} \sim 10^9$ M$_\odot$; or $\sim 18$ $e$-folding times for a stellar mass BH seed. This is barely allowed for the traditional Salpeter times of $\sim 45$ ($\varepsilon/0.1$) Myr, assuming an Eddington-limited accretion.

For a Schwarzschild BH, the standard thin disc radiative efficiency is $\varepsilon \approx 0.06$, and a seed BH of $\sim 10^2$ M$_\odot$, e.g., a remnant of Population III stars, can grow up to a few billion solar masses by $z \sim 6$. For spinning BHs, however, radiative efficiencies can approach 20% (magnetized disc accretion; see Gammie et al. 2004, Krolik et al. 2005) or even 30% (unmagnetized disc accretion, Thorne 1974). With such a high efficiency, it takes longer than $\sim 1 - 2$ Gyr for the seeds to join the supermassive variety. Accretion rates above the Eddington limit, or more massive seeds might ease the strict requirements that $z \sim 6$ quasars pose to SMBH growth (Fig. 2).

SMBHs can grow through accretion and merging, but they must start from some lower mass BH seeds. The question of the possible formation mechanisms for SMBHs was already addressed in the classical review paper by Rees (1978). Cast in the present parlance, there are essentially four generic possibilities: (1) stellar mass BHs produced as endpoints of the massive star evolution; (2) direct gravitational collapse of dense protogalactic cores; (3) gravitational runaway collapse of dense star clusters; (4) primordial BH remnants from the big bang. The only mechanism actually known to occur in nature is (1), but (2) and (3) are certainly possible, and while (4) seems a bit of a *deus ex machina*, and there is no evidence for the existence of such primordial BH population, it could reflect some interesting fundamental physics. Of course, a combination of these seeding mechanisms may be at play. We will address them in turn.

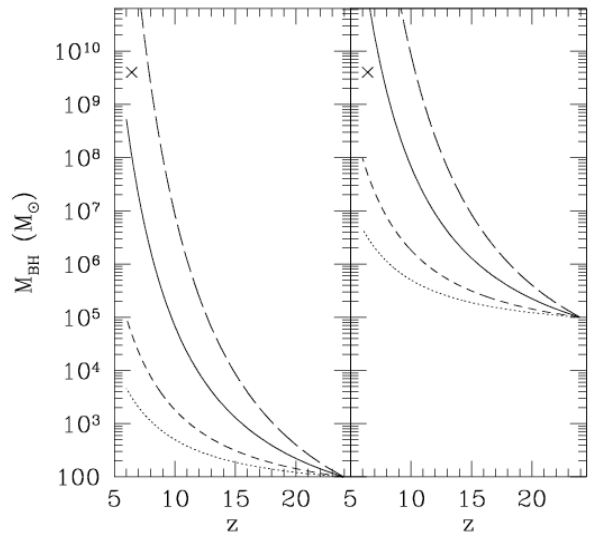

**Figure 2.** Growth of a BH mass under different assumption for the initial mass efficiency, assuming Eddington limited accretion. Left panel: Pop III remnants seeds ($M_{BH} \approx 10^2 \, M_\odot$). Right panel: direct collapse seeds ($M_{BH} \approx 10^4 - 10^6 \, M_\odot$). Long-dashed line: $\varepsilon \approx 0.06$ (Schwarzschild BH); solid line: $\varepsilon \approx 0.1$; short –dashed line: $\varepsilon \approx 0.2$; dotted line: $\varepsilon \approx 0.3$.

### 3.1 Stellar-Mass BH Seeds from Population III Stars

The first stars, which formed in dark matter (DM) minihalos of mass $\sim 10^6 \, M_\odot$ at redshifts of $z \sim 20$, were likely very massive, having characteristic masses of the order of $\sim 100$ $M_\odot$ (Bromm et al. 1999, 2002, Nakamura & Umemura 2001, Abel et al. 2002, Yoshida et al. 2006). This prediction ultimately relies on the absence of efficient cooling agents in the primordial, pure H/He gas, typically having temperatures below $\sim 10^4$ K, the threshold for the onset of atomic hydrogen cooling (Bromm & Loeb 2004, Glover 2005). Under such conditions, the only viable coolant is molecular hydrogen ($H_2$), and possibly HD (Johnson & Bromm 2006). These massive Population III stars would have radiated at temperatures of $\sim 10^5$ K (Bromm et al. 2001), generating enough ionizing photons to completely ionize the minihalos in which they were formed and to contribute to the reionization of the universe (Barkana & Loeb 2001, Alvarez et al. 2006). Those Pop III stars with masses $40 \, M_\odot < M_* < 140 \, M_\odot$ or $M_* > 260 \, M_\odot$ are predicted to have collapsed to form BHs directly, possibly providing the seeds for the first quasars (Madau & Rees 2001, Heger et al. 2003, Madau et al. 2004, Ricotti & Ostriker 2004, Kuhlen & Madau 2005) although more massive black holes may have been formed after the epoch of the first stars in DM halos with virial temperatures of $\sim 10^4$ K (Bromm & Loeb 2003, Spaans & Silk 2006, Begelman et al. 2006).

The growth of the first BHs must have been rapid enough to account for the powerful QSOs observed at $z > 6$ (Haiman & Loeb 2001, Volonteri & Rees 2005). It is not obvious how such vigorous accretion could have taken place, at least early on, just after the formation of the Pop III remnant BH. It has been demonstrated that the radiation from the first stars heats and evacuates the gas residing within the $\sim 10^6 \, M_\odot$ minihalos in which they were born (Whalen et al. 2004, Kitayama et al. 2004, Alvarez et al. 2006). To enable efficient accretion, the baryonic mass around these BHs must therefore have been replenished. In the course of hierarchical structure formation, this continued accretion of matter is naturally accomplished through mergers of the BH's parent halo with neighboring halos (Ricotti & Ostriker 2004, Kuhlen & Madau 2005). The crucial question is how quickly this merger-driven inflow of material could have begun to feed the seed BH, given that it is surrounded by a high-pressure, relic H II region. Recently, we have gained first hints from numerical simulations that take into account the relevant chemical and thermal evolution in the vicinity of the Pop III remnants (Johnson & Bromm 2007, Pelupessy et al. 2007).

A black hole with an initial mass of $\sim 100 \, M_\odot$ must have accreted with an average rate of $> 1 \, M_\odot \, \mathrm{yr}^{-1}$ to attain masses of $\sim 10^9 \, M_\odot$, close to what has been inferred for the black holes that power luminous QSOs at $z > 6$. It is instructive to estimate the accretion rate of a BH that begins with a mass $m_i$ and the resulting BH mass $m$ as a function of time $t$ since its formation for a given density and temperature of accreted gas (Johnson & Bromm 2007). Assuming that the black hole accretes gas from a cloud with uniform density and temperature and with dimensions much greater than the accretion radius of the black hole, $r_{acc} \sim Gm/c_s^2$, where $c_s$ is the speed of sound. In this case, one can estimate the accretion rate as (Bondi 1952):

$$\frac{dm}{dt} \sim \frac{2\pi (Gm)^2 m_\mathrm{H} n}{c_\mathrm{s}^3}$$

where $c_s$ is the speed of sound of the gas at infinity, $m_H$ is the mass of the hydrogen atom, and $n$ is the number density of the gas at infinity. Assuming that a fraction $\epsilon$ of the accreted mass is converted into energy and radiated away, integrating from $t = 0$ and $m = m_i$ yields a time dependent mass of the BH of

$$m \sim \left( \frac{1}{m_\mathrm{i}} - \frac{2\pi (1-\epsilon) m_\mathrm{H}^{\frac{5}{2}} G^2 n t}{(3 k_\mathrm{B} T)^{\frac{3}{2}}} \right)^{-1}$$

where $T$ is the temperature of the gas at infinity and $c_s{}^2 = 3 \, k_B T / m_H$.

To estimate the growth rates of the first BHs formed by the direct collapse of Pop III stars, one can use the densities and temperatures of accreting gas derived from recent simulations of the evolution of the relic H II region that would have surrounded these first $\sim 100 \, M_\odot$ BHs (Johnson & Bromm 2007). For the purpose of deriving a simple estimate, one can neglect the effects of radiative feedback on the accretion rate of the BH, although it has been shown that at high accretion rates the radiation emitted from quasars can inhibit further gas infall (Di Mateo et al. 2005, Springel et al. 2005a). This assumption is partially justified by the low accretion rates and associated luminosities encountered in the simulations (Johnson & Bromm 2007), although it is dependent on the uncertain efficiency with which the accreting gas may absorb the emitted energy. The estimated accretion rates are thus upper limits.

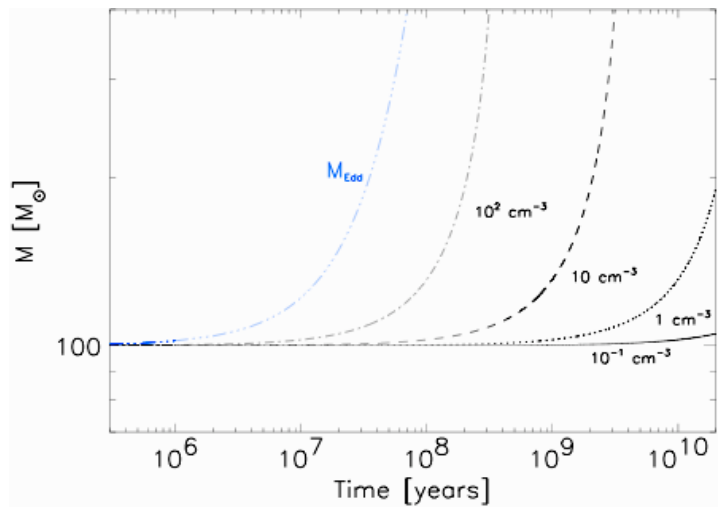

**Figure 3.** Early accretion on Pop III BH seeds (adopted from Johnson & Bromm 2007). The mass of an initially 100 $M_\odot$ BH as a function of time, assuming that the BH accretes gas at a temperature of 200 K and at a constant density. The solid, dashed, dotted, and dotted-dashed lines show cases with gas densities of 0.1 cm⁻³, 1 cm⁻³, 10 cm⁻³, and 100 cm⁻³, respectively. The triple dot-dashed line shows the mass of the BH as a function of time, assuming that it accretes at the Eddington limit.

Fig. 3 shows the mass of an initially 100 $M_\odot$ BH as a function of time, for various gas densities and assuming a temperature for the accreting gas of 200 K, which is the temperature of the highest density gas found in the simulation of the relic H II region (Johnson & Bromm 2007). In the case of Eddington-limited accretion, the BH attains a mass of ~$10^9$ $M_\odot$ in ~800 Myr, indicating that such a BH could marginally grow to become a SMBH by $z \sim 6$, and power the observed SDSS QSOs. However, the simulations are finding extremely low densities, $n < 1$ cm⁻³, in the vicinity of the remnant BH that cannot be replenished for at least ~ $10^8$ yr (Johnson & Bromm 2007). Indeed, for densities lower than ~ 600 cm⁻³, or gas temperatures higher than ~ 200 K, an initially ~ 100 $M_\odot$ BH likely cannot grow fast enough to power the high-$z$ QSOs. BHs of initially higher masses could accrete at higher rates, since the accretion rate varies as $m^2$. An initially 500 $M_\odot$ BH would require accreting gas densities > $10^2$ cm⁻³ at temperatures < 200 K in order to accrete at the Eddington limit. Continuously accreting at or near the Eddington limit, and starting shortly after its formation by the collapse of a Pop III star at $z > 20$, such a BH could reach a mass of $10^9$ $M_\odot$ within ~700 Myr, and thus could be the progenitor of the SMBHs at $z \sim 6$.  However, numerical simulations (Johnson & Bromm 2007) indicate that gas densities in the vicinity of the remnant BH remain low for a Hubble time ($\sim 10^8$ yr at $z \sim 20$), even taking into account the merger-driven inflow of material in the course of hierarchical structure formation.  Therefore, stellar seed scenarios might face a challenging early bottleneck in terms of getting efficient accretion going to allow growth to the inferred quasar SMBHs at $z \sim 6$.

This is relevant for another important issue, the nature of sources which reionized the universe (Barkana & Loeb 2001), and specifically the relative contributions from stars and quasars.  Reionization likely was an extended process that already began at $z > 11$, reaching back to the epoch of the first stars (see, e.g., Greif & Bromm 2006).  It has been suggested that microquasar activity, powered by accretion onto Pop III BH remnants, could have played an important role in the first stages of reionization (Madau et al. 2004, Ricotti & Ostriker 2004, Kuhlen & Madau 2005, etc.)  Different from stellar sources, which only produce ionizing radiation for a few Myr, microquasars could shine for much longer times, provided an efficient accretion.  At least initially, the conditions for accretion at the Eddington limit might not be met, and the miniquasars fueled by accretion onto the Pop III remnant BHs would not emit radiation at the Eddington limit until well after the formation of the BHs.  If accretion onto miniquasars is inefficient, the possible generation of part of the observed near-IR background excess by Pop III stars (Santos et al. 2002, Kashlinsky et al. 2005) may not necessarily imply an overproduction of X-rays emitted by miniquasars fueled by accretion onto Pop III remnant BHs (Madau & Silk 2005).  However, copious X-ray production could result from accretion of gas from a binary companion (Belczynski et al. 2004) even if the binary system resides in a low density environment.  Future work will have to include more detailed, fully cosmological simulations with a prescription for radiative feedback from Pop III remnant black holes to self-consistently address these issues.

### 3.2 Direct Formation of SMBHs

Given the difficulties posed by the growth of stellar mass BH seeds, one possibility is that black holes form already massive, in the cores of proto-galaxies, from a collapsing gas cloud (Haehnelt & Rees 1993, Loeb & Rasio 1994, Eisenstein & Loeb 1995, Bromm & Loeb 2003, Koushiappas, Bullock & Dekel 2004, Begelman, Volonteri & Rees 2006, Lodato & Natarajan 2006, Begelman 2007, etc.).

In halos where cooling is efficient, the gaseous component can cool and contract, until further collapse is halted by rotational support.  Halos, and their baryonic cores, possess angular momentum, believed to have been acquired by tidal torques due to interactions with neighboring halos.  The halos angular momentum, $J$, can be related to the so-called spin parameter $\lambda \equiv J |E|^{1/2} / G\, M_{\mathrm{h}}^{5/2}$, where $E$ and $M_{\mathrm{h}}$ are the total energy and mass of the halo.  The imparted transverse velocity, however, is only $\sim 5\%$ of what is needed for full rotational support.  Cold gas can therefore infall and condense toward the center until angular momentum becomes important. Let us assume that a fraction $f_d$ of the gas can cool, and settles into an, e.g., isothermal, exponential disc embedded in a dark matter halo described by a Navarro, Frenk & White (1997) density profile. Under these assumptions, the scale radius of the gaseous disc is a fraction $\sim \lambda$ of the virial radius of the halo. The tidally induced angular momentum of the gas would therefore be enough to provide centrifugal support at a distance of a few tens of pc from the center, where a self-gravitating disc forms. Additional mechanisms inducing transport of angular momentum are needed to condense further the gas. The removal of angular momentum must also occur on timescales shorter than star formation, to avoid excessive consumption of gas. The loss of angular momentum can be driven either by (turbulent) viscosity or by global dynamical instabilities, such as the "bars-within-bars" mechanism (Shlosman, Frank & Begelman 1989). The gas can thus condense to form a central massive object (Fig. 4).

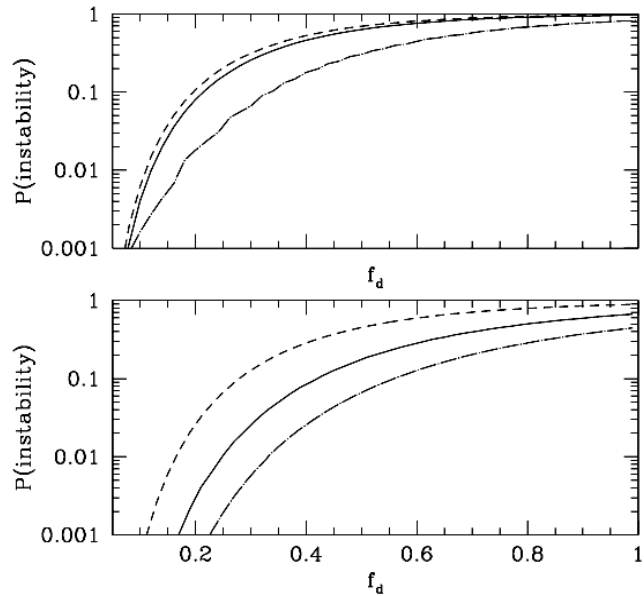

**Figure 4.** Fraction of metal free halos with virial temperature $10^4$ K at $z = 1$ 8 which are prone to viscous (lower panel, see Lodato & Natarajan 2006) or gravitational (upper panel, see Begelman, Volonteri & Rees 2006) instabilities, which can lead to the conditions suitable for BH formation. The fraction of unstable systems is shown as a function of the fraction of gas participating to the collapse process. Different line styles refer to different choices of stability parameters.

The mass of the seeds predicted by different models vary, but typically they are in the range $M_{BH} \sim 10^4 - 10^6 \, M_\odot$. Models based on angular momentum removal via viscosity typically postulate that the infalling gas collapses to form a supermassive star ($>10^4 \, M_\odot$), which eventually becomes subject to post-Newtonian gravitational instability and forms a seed BH. Begelman, Volonteri & Rees (2006) follow instead the collapse of gas prone to the "bars-within-bars" instability. The very rapid accumulation of gas ends up creating a low-entropy star-like configuration where a small BH ($\sim 10 - 20 \, M_\odot$) forms in the core, via neutrino cooling. The BH then grows very rapidly, at super-Eddington rates, by accreting the surrounding envelope.

Bromm & Loeb (2003) simulated the collapse of the gas in metal-free halos with low to average angular momentum, and noticed that the efficient collection of gas in the center of a halo probably occurs only under metal-free conditions where the formation of $H_2$ is inhibited, for instance due to an intergalactic UV background. In the presence of metals, or of $H_2$, the gas cools to low temperatures and fragmentation and star formation is favored over BH formation. If only atomic hydrogen is present, however, the gas does not cool much below $10^4$ K and star formation is suppressed. Begelman, Volonteri & Rees (2006) have shown that if MBH formation proceeds until late times, the total mass density in MBH seeds can be exceedingly large (e.g., larger than the mass density in SMBHs observed at the present time; Yu & Tremaine 2002, Marconi et al. 2004). MBH formation is likely to terminate when the gas in halos reaches the critical metallicity threshold for fragmentation (Santoro & Shull 2006, Frebel et al. 2007).

### 3.3 Relativistic and Runaway Collapse of Dense Star Clusters

Gravitational collapse of dense star clusters provides another astrophysically viable mechanism for generation of IMBHs, since we do know that dense star clusters do exist, they are often produced in regions of active star formation (and we can see many recent examples in nearby galaxies, including the Magellanic Clouds). Moreover, it has been known for a long time that the dynamical evolution of any star cluster, with stars of different individual masses, will be characterized by a contracting core and an expanding envelope, leading to a mass-segregation instability (Spitzer 1969). Such a phenomenon brings slowly this stellar system towards the gravothermal catastrophe and core collapse, which can occur even in the absence of the mass segregation, as long as there is a sufficient density contrast within the cluster (Antonov 1962, Hénon 1961, 1965). Such an evolution is directly related to high stellar number densities, which enhence star collision rates (Heggie 1975). However, while cluster cores may and do collapse, they can be stabilized by binaries and not continue all the way to a BH formation.

Simulations of cluster collapse (as a cloud of stars as mass points) into an IMBH have been studied, e.g., by Shapiro & Teukolsky (1985,1986), Kochanek et al. (1987), Shapiro (2004, and references therein). A somewhat different mechanism is formation of BHs via runaway merging of the most massive stars, especially in young clusters. This has been studied in modern simulations, e.g., in clusters with relaxation times $t_{rlx} \leq 25$ Myr, by Portegies Zwart & McMillan (2002). In the case of old star clusters, growth of IMBHs in globular clusters (GCs) through a series of binary mergers has been studied by Gültekin et al. (2004), Kawakatu & Umemura (2005), and O'Leary et al. (2006), among others. Both of these mechanisms could have produced IMBH remnants in present-day globular clusters, which we discuss below, and could have also produced seed IMBHs in young galaxy cores at high redshifts. Since the expected masses of such cluster collapse BH seeds would be in the range $\sim 10^2 - 10^4$ $M_\odot$, thus alleviating the early growth time scale problems, they would be very attractive as the progenitors of SMBHs observed in the high-redshift QSOs.

### 3.4 Primordial BH Seeds

Finally, it is possible that a suitable population of BH seeds may be generated in the early stages of the big bang (Hawking 1971). Excellent reviews of the subject include, e.g., Carr (2003, 2005ab), or Khlopov (2008). In the context of this review, relevan modern treatments include, e.g., Ricotti & Ostriker (2004), Ricotti et al. (2005, 2008), Mack et al. (2007), or Ricotti (2007).

Whereas there is no compelling evidence for such a population of primordial BHs, the limits on a population of such BHs with masses ~ stellar mass or larger are not very good, and it would be worthwhile to explore their possible observable manifestations which are not yet excluded by observations. Discovery of such objects could have a profound impact on our inderstanding of the early universe, and the physics of gravity.

## 4    The Early Growth of SMBHs, and Relation to Galaxy Formation

Once formed, BHs will grow through accretion and hierarchical merging. We note that the accretion generates observable elctromagnetic signatures of this process, while merging simply rearranges the distribution of mass already present in the BHs.

As they grow, SMBHs would manifest themselves as AGN, and affect their host galaxies and large-scale environments. Some of these issues are also reviewed by Springel elsewhere in this volume.

### 4.1 Dissipative Merging and Starburst Connection: Numerical Modeling

A prerequisite for the rapid growth of SMBHs by gas accretion is an efficient QSO fueling mechanism, capable of delivering copious amounts of gas to the very central region of a galaxy. Major galaxy mergers are a natural candidate to facilitate the required large-scale gaseous inflows, because the gravitational tidal forces during galaxy collisions can efficiently extract angular momentum from gas in the interstellar medium of disk galaxies, and drive it to the center. In particular, it is commonly believed that the central gas increase resulting from mergers gives rise to powerful nuclear starbursts, a scenario that has been studied in substantial detail using hydrodynamical simulation models of galaxy mergers (Barnes & Hernquist 1992, Mihos & Hernquist 1994). By the same token, the nuclear inflow of gas triggered by galaxy interactions may feed an embedded SMBH, and can thereby cause QSO activity. Indeed, semi-analytic models for galaxy and BH growth can successfully reproduce basic properties of the QSO population and its evolution based on the assumption that QSOs are triggered by major galaxy mergers (Kauffmann & Haehnelt 2000 ).

Recently, Di Matteo et al. (2005), Springel et al. (2005a), and Li et al. (2007) have extended the numerical methods used for hydrodynamic simulations of galaxy mergers to also follow the growth of central SMBHs by gas accretion, besides tracking radiative cooling processes, star formation, and the collisional dynamics of dark matter and stars. Even though the physical details of the fueling process close to the SMBH remain comparatively poorly understood and cannot directly be treated and resolved in simulations of whole galaxies, the calculations provide accurate estimates of the gas inflow rates into the central regions of merging galaxies.

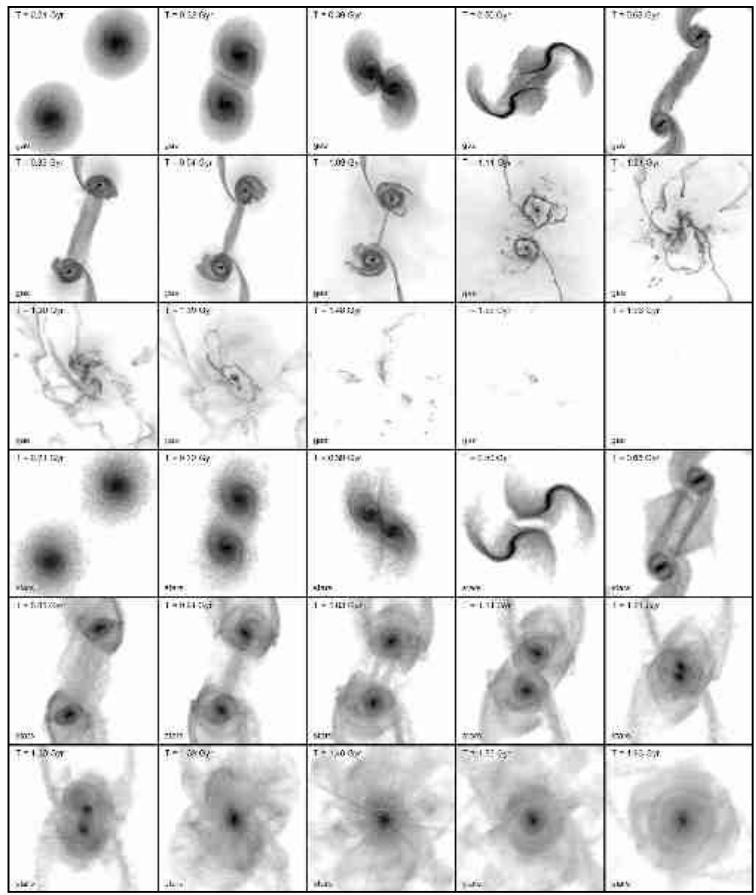

(Version of the paper with a full resolution figure is available at http://www.astro.caltech.edu/~george/mg11/ )

**Figure 5.** Time evolution of a disk galaxy merger with quasar growth and associated energy feedback. The images in the top three rows show the projected gas density, color-coded by local gas temperature. The bottom three rows give the corresponding projected stellar density, color-coded by the mean age of the stellar population. The luminous spikes in these images indicate the quasar luminosity (while the spikes are an artists conception, their total brightness is proportional to the quasar luminosity). From Di Mateo et al. (2005).

Under the assumption that this large-scale inflow rate is ultimately determining the extent of the possible BH growth, the accretion rate can be estimated based on a simple Bondi rate prescription. The estimated accretion rate was however not allowed to be higher than the Eddington rate, as the radiation field is not dynamically followed in the simulations. A further assumption made in their models is that a small fraction of 5% of the radiative luminosity associated with the accretion (which is taken to be 10% of the accreted rest mass energy) couples thermally to the local gas in the surroundings of the BH. This provides a feedback mechanism for the accretion, which can therefore exhibit a self-regulated behavior.

Figure 5 shows a visualization of a typical time evolution of a merger of two equal mass disk galaxies on a prograde orbit calculated with this model. The series of panels in the top three rows gives the evolution of the projected gas density, while the bottom rows show the stellar density, as labelled. The tidal forces during the first encounter of the galaxies distort the disks into a pair of strongly barred galaxies. A phase lag between the stellar and gaseous bar efficiently extracts angular moment from the gas, such that a substantial amount of gas flows into the nuclei, where it fuels both a central starburst and QSO accretion. At first, the BHs are small, and despite accreting rapidly, their feedback energy has an insignificant effect on the surrounding gas. However, eventually the BHs become sufficiently massive during the exponential, Eddington-limited growth phases that are caused by the rapid gas fueling that their own feedback terminates the further growth by driving a pressure-driven outflow. This happens both at an intermediate phase when the galaxies undergo a first starburst triggered by the initial encounter, and a second time when the galaxies finally coalesce. The gas-fueling during the final coalescence leads to further substantial growth of the remnant BH, which shines as a luminous QSO when the obscuring gas column becomes small enough. The QSO activity is terminated on a short timescale once the accretion luminosity becomes sufficient to unbind the gas from the central potential well, and to drive a powerful outflow. As a result, a gas-poor "dead" elliptical galaxy with very low star formation rate is left behind.

During the final coalescence, the BHs come quite close to each other with a small relative velocity, such that they plausible form a SMBH binary. The numerical model simply merges the black holes in this case, i.e. it is assumed that binaries of supermassive black holes merge efficiently on a short time-scale. A potential recoil from the emission of gravitational radiation was neglected.

An important consequence of the self-limited growth of BHs is that this mechanism can provide an explanation for the observed correlation between BH mass and stellar velocity dispersion of the hosting bulge. Numerical results for the latter are compared to observational data in Figure 6, showing that the numerical models are capable of reproducing the observed relationship quite well. In larger halos, the BH needs to grow to larger masses before its energy feedback can expel and unbind the gas from the central gravitational potential well. The final BH masses obtained in the merger simulations are also found to be quite insensitive to the gas fractions and structural properties of the progenitor galaxies. The simulation models hence lend strong support to earlier theoretical suggestions that a self-limited growth of BHs might be the reason for the scaling relations observed for the SMBH population (Silk & Rees 1998, Wyithe & Loeb 2003). As galaxy mergers induce a common formation of SMBHs and their hosting spheroids, they naturally provide for a link between QSO and galaxy formation. Based on this link, Hopkins et al. (2006ab) have conjectured a comprehensive model for a merger-driven origin of starbursts, QSOs, galaxy spheroids, SMBHs, and the cosmic x-ray back background, that very successfully matches a diverse set of observational data sets.

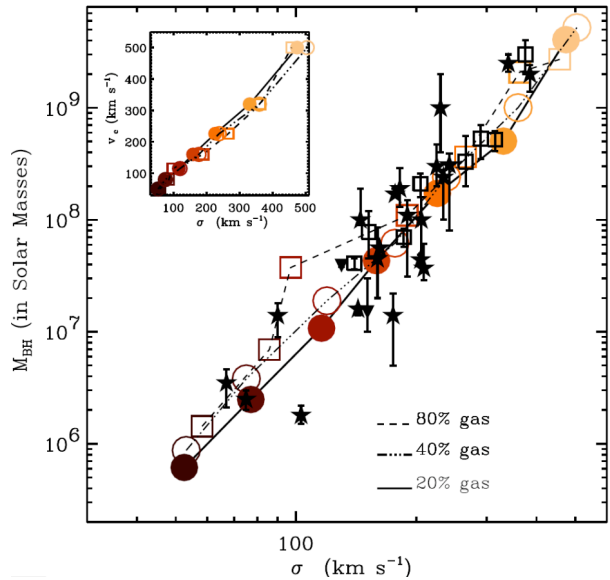

**Figure 6.** Relation between BH mass and stellar velocity dispersion in the merger remnants of simulations of disk galaxy mergers (Di Matteo et al. 2005). The solid circles give results for the final remnants of six merger simulations of galaxies with disk gas fraction of 20%, but different total mass, parameterized by virial velocities of $V_{vir}$ = 50, 80, 160, 320, and 500 km/s. Open circles and open squares with the same color give results for gas fractions of 40% and 80%, respectively. Black symbols show observational data for the masses of SMBHs and the velocity dispersions of their host bulges (Tremaine et al. 2002, Ferrarese & Ford 2005).

Another interesting consequence is that quasar feedback can lead to a rapid shut-down of star formation, and to the expulsion of significant amounts of gas from the formed remnant galaxy. Especially in gas-rich mergers of massive galaxies, this substantially reduces the residual star formation rate once the merger has been completed, allowing the merger remnant to develop stellar colors that are as red as observed for massive ellipticals (Springel et al. 2005b, Sijacki et al. 2007).

We note that the QSO feedback discussed here needs to be distinguished from the notion of "radio-mode" feedback that has recently been invoked to explain the absence of strong cooling flows in clusters of galaxies (Croton et al. 2006, Sijacki & Springel 2006). It is believed that the black holes in such radio-loud systems are in a low accretion rate state (with Eddington ratios below 0.01) that is characterized by a radiatively inefficient accretion flow that is presumably quite different in structure than the accretion flows during QSO activity. This radio-mode becomes important at low redshift, and in massive systems, while the QSO activity shows a strong decline towards the present epoch, in line with the reduction of the merger rate of galaxies.

### 4.2 Observable Tests: The Evolving QSO Luminosity Function and Environments

Another interesting result of the merger simulations discussed above is that they predict a specific dependence of the lifetime of QSOs on their peak luminosity. As Hopkins et al. (2005ab) showed, more luminous QSOs exhibit systematically different lightcurves, and in particular, shorter lifetimes. This is quite different from the standard assumptions most often made in theoretical models for the QSO population, where the lightcurve is either modeled with a constant luminosity for some finite lifetime ("lightbulb" model), or where the luminosity grows exponentially for some time at the Eddington rate, and then turns off abruptly ("exponential model"). More general lightcurves for a QSO of a peak luminosity $L_{peak}$ can be characterized by the function (d$t$ / d $log\,L$), i.e., the fractional time a QSO spends per logarithmic luminosity interval. The galaxy merger simulations can be used to quantitatively determine this function (Hopkins et al. 2005a). It is found that QSOs spend significant amounts of time at sub-Eddington accretion rates during BH growth, with a distribution that depends on the final BH mass reached. The QSO luminosity function (QLF) can be written as a convolution of the lifetime function with the rate d$n(L_{peak})$/d$t$ at which quasars of a certain peak luminosity $L_{peak}$ are formed, viz.,

$$\phi(L) \propto \int \frac{\mathrm{d}t}{\mathrm{d}\log L} \dot{n}(L_{\mathrm{peak}}) \,\mathrm{d}\log L_{\mathrm{peak}}$$

Using the QSO *lifetime* function provided by the simulations together with the observed QLF, one can also infer the formation rate of QSOs by deconvolution from this equation.

In Figure 7, the observed hard X-ray QLF from Ueda et al. (2003) is compared to the predictions of Hopkins et al. (2006) based on their model for QSO lifetimes and evolution. For luminosity-dependent lifetimes, the distribution function of the QSO formation rate has a very different shape compared with the QLF itself, unlike in light-bulb or exponential lightcurve models, where these shapes are identical. In particular, the population of faint QSOs is not really composed of intrinsically low-luminosity QSOs when luminosity dependent lifetimes are assumed. Rather, they are QSOs either on the way in or on their way out of their peak activity. (Recall that they they spend most of their time in a state of low luminosity.) In contrast, the bright end of the QLF directly reflects the shape of the underlying formation rate of SMBHs.

An important implication of the above is that QSOs are predicted to occupy a relatively narrow mass range of host dark matter halos. On average, bright and faint QSOs reside in similar host halos. As a result, the clustering amplitude of QSOs is expected to only show a weak dependence on luminosity, unlike in alternative models for QSO lightcurves (Lidz at al. 2006). Observational constraints for the clustering strength (or bias) as a function of luminosity provide therefore a powerful way to directly constrain this scenario, and hence to inform different models for QSO feedback. Interestingly, some current observational constraints for the luminosity dependence of the QSO correlation function (Croom et al. 2005) and galaxy-QSO cross-correlation function (Adelberger & Steidel 2005) appear to favor the picture of luminosity dependent lifetimes, but the large error bars and limited dynamic range in luminosity of the present observational samples preclude strong conclusions at this point. On the other hand, there is a growing evidence for strong biasing of high-redshift QSO samples (Djorgovski et al. 2003, Stiavelli et al. 2005, Shen et al. 2007, etc.).

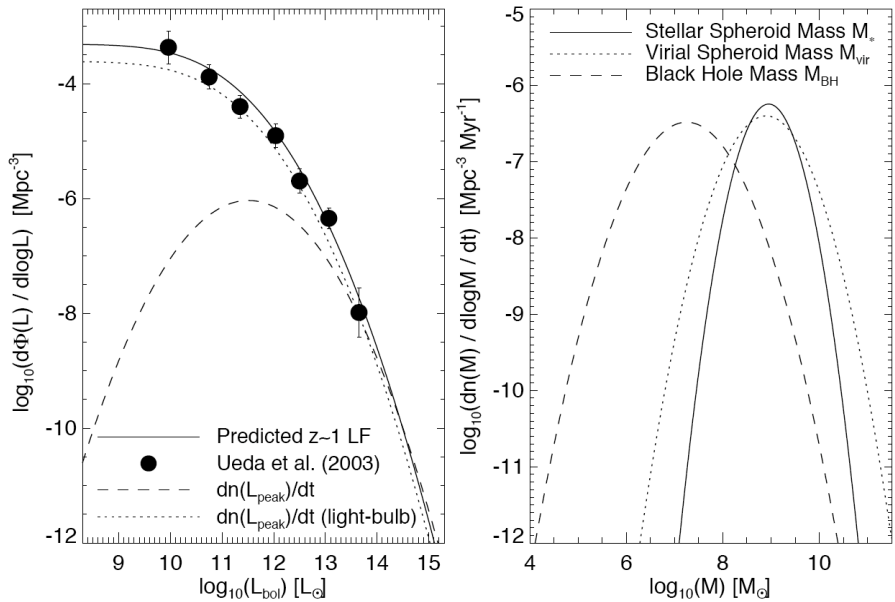

**Figure 7.** Left: The observed X-ray QLF (dots) comared to the predictions of Hopkins et al. (2006) model (solid line), and the inferred formation rate of QSOs (dashed), and the light-bulb or exponential lightcurve models (dotted). Right: The inferred formation rate of SMBHs/QSOs as a function of BH mass (dashed), halo mass (dotted), or stellar mass of the corresponding spheroidal galaxy (solid).

Another important observational test of the QSO feedback scenario would be the detection of QSO outflows, and in particular, in an accurate measurement of their kinetic luminosities. While it is now widely recognized that most QSOs indeed show some signs of outflows (Richards 2006, Gabel et al. 2006, Arav et al. 2007), large uncertainties e.g. in the covering factors make it difficult to estimate the energy carried by these winds.

*4.3 SMBH Mergers and Gravitational Waves*

SMBHs should evolve along with the hierarchy of mergers of their host galaxies, and share their fate, growing by reassembly, along with accretion. Dynamical friction drives efficiently mergers of galaxies, and accelerates merging of their central BHs. The efficiency of dynamical friction decays when the BHs get close and form a binary, when the binary separation is around a few tenths of pc (for $M_{BH} \sim 10^4 - 10^6 \, M_\odot$). Emission of gravitational waves becomes efficient at binary separations about two orders of magnitude smaller. Evolution of BH binaries can also have profound dynamical effects on the cores of their host galaxies; for a review, see, e.g., Merritt (2006), and references therein.

In gas-poor systems, the subsequent evolution of the binary, while gravitational radiation emission is negligible, may be largely determined by three-body interactions with background stars (Begelman, Blandford & Rees 1980), by capturing the stars that pass within a distance of the order of the binary semi-major axis and ejecting them at much higher velocities (Quinlan 1996, Milosavljevic & Merritt 2001, Sesana, Haardt & Madau 2006). Dark matter particles will be ejected by decaying binaries in the same way as the stars, i.e. through the gravitational slingshot. In minihalos a numerous population of low-mass stars may be present if the IMF were bimodal, with a second peak at $1 - 2$ $M_\odot$, as suggested by Nakamura & Umemura (2001). Otherwise the binary will be losing orbital energy to the dark matter background. The hardening of the binary modifies the density profile, removing mass interior to the binary orbit, depleting the galaxy core of stars and dark matter, and slowing down further decay. Let us assume (e.g.,Volonteri, Haardt & Madau 2003) that the removal of matter creates a core of constant density, eroding a preexisting isothermal cusp with $\rho \propto r^{-2}$, and that BH fall on the same relation with the velocity dispersion of the host as they do today. The timescale for merging of an equal mass binary can be much longer than the Hubble time then.

In gas rich high redshift halos, however, the orbital evolution of the central SMBH is likely dominated by dynamical friction against the surrounding gaseous medium. The available simulations (Mayer et al. 2006, 2007, Dotti et al. 2007, Escala et al. 2004, 2005) show that the binary can shrink to about parsec or slightly subparsec scale by dynamical friction against the gas, depending on the gas thermodynamics. The interaction between a SMBH binary and an accretion disc can also lead to a very efficient transport of angular momentum, and drive the secondary BH to the regime where emission of gravitational radiation dominates on short timescales, comparable to the viscous timescale (Armitage & Natarajan 2005, Gould & Rix 2000).

The final stages of the BH mergers is dominated by emission of gravitational waves: the merging history of SMBHs along the hierarchical build-up of cosmic structures leaves a unique imprint on the gravitational waves sky. The detection of the emission of gravitational waves from merging BH seeds at mHz frequency using future gravitational wave missions, such as LISA, is likely the most promising way of disentangling the very early evolution of BH seeds, with masses below $\sim 10^6 - 10^7 \, M_\odot$ at $z > 10$.

Direct electromagnetic observations of high-redshift miniquasars powered by BHs with mass of a few thousands $M_\odot$ are beyond the reach of current, or even foseeable ground-based and space-based observatories. LISA, however, in principle is sensible to gravitational waves from binary BHs with masses in the range $\sim 10^4 - 10^6\ M_\odot$ at virtually any redshift of interest.

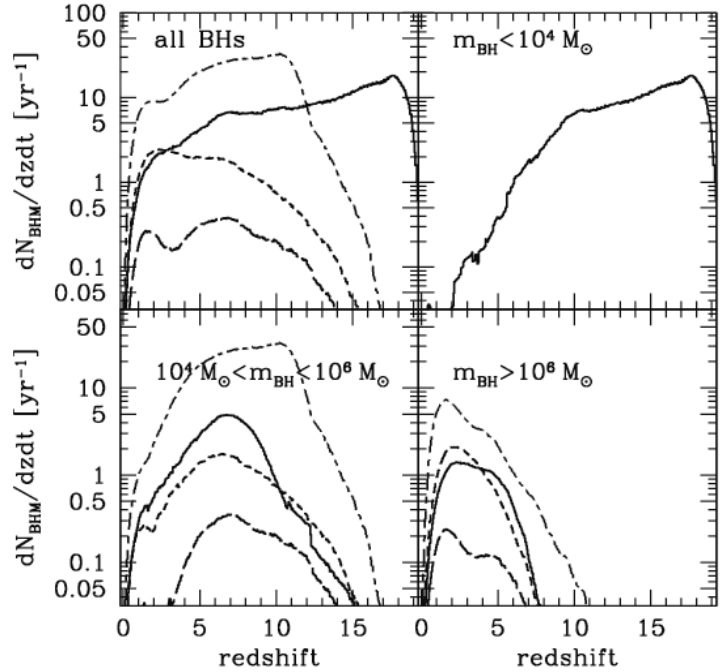

**Figure 8.** Number of BH mergers per observed year at $z = 0$, per unit redshift, in different BH binary mass intervals. Solid lines: remnants of Pop. III stars (Volonteri, Haardt & Madau 2003), short-long dashed lines: Koushiappas, Bullock & Dekel (2004) model; short-dashed and long-dashed lines: Begelman, Volonteri & Rees (2006) model, assuming two different BH formation efficiencies. From Sesana, Volonteri & Haardt (2006).

Let us consider two main scenarios for SMBH formation, namely, one where seeds are light ($M_{BH} \sim 10^2\ M_\odot$, remnants of Population III stars), and one where seeds are heavy ($M_{BH} \sim 10^4 - 10^6\ M_\odot$, direct collapse). Detection of several hundreds merging events in a 3 year LISA mission will be the sign of a heavy seed scenario with efficient formation of BH seeds in a large fraction of high redshift halos (Koushiappas, Bullock & Dekel 2004). On the other extreme, a low event rate, about a few tens in 3 years, is typical of scenarios where either the seeds are light, and many coalescences do not fall into the LISA band, or seeds are massive, but rare (e.g., Begelman, Volonteri & Rees 2006, Lodato & Natarajan 2006). In this case a decisive diagnostic is provided by the mass ratios of binary coalescences. Direct collapse models predict that most of the detected events involve equal mass binaries. A large fraction of observable coalescences, in fact, involve BHs at $z > 10$, when seeds had little time to accrete much mass yet. In scenarios based on Population III remnants, $z > 10$ mergers involve BHs with mass below the LISA threshold. The detectable events happen at later times, when MBHs have already experienced a great deal of mass growth.

### 4.4 Possible Electromagnetic Signatures of Cosmological BH Mergers

The gravitational wave astronomy is still in its infancy, and it would be very interesting to find if mergers of BHs during the hierarchical assembly of galaxies also provide some observable electromagnetic signature. Detections of such events may precede, and would certainly greatly enhance the scientific returns from the gravitational wave detections (see, e.g., Stubbs 2008, and references therein).

At first, the subject seems unpromising, since the BH binaries can very effectively clear out their immediate surroundings prior to the merger. However, these merger environments could be messy. Shocks in the leftover, hollowed-out accretion disk from a pre-merger activity could create a transient X-ray source (Milosavljevic & Phinney 2005). Also, if either or both of the merging SMBHs were accreting prior to their coalescence, one could look for a QSO which disappears prior to the gravitational wave signal (~ preglow); or conversely, if the product SMBH starts accreting any baryons in the presumably messy environment of the merger, a QSO (~ afterglow) would appear following the event (Dotti et al. 2006). Detections of any QSOs positionally coincident with the BH-merger type gravitational wave events would provide valuable information about their physics (Kocsis et al. 2007, 2008).

As merging BH binaries are expected to experience recoils from gravitational wave emission, encounters of ambient gas with the leftover accretion disks around the SMBH remnant could generate shocks which would manifest as infrared or X-ray sources (Bonning et al. 2007, Shields & Bonning 2008, Schnittman & Krolik 2008, Lippai et al. 2008). It is probably fair to say that the exact manifestations of such phenomena would depend very strongly on the structure and extent of the baryonic environments of BH mergers.

It is thus not hopeless to search for transient electromagnetic phenomena associated with BH mergers and resulting gravitational wave signals; and surprises and unexpected new phenomenology are certainly possible. Finding any such transients in the ongoing or upcoming synoptic sky surveys (see, e.g., Djorgovski et al. 2008 for a current example, and LSST, http://www.lsst.org, for a future one) would be a very exciting discovery.

## 5 IMBH Remnants at Low Redshifts?

As the SMBHs grow from lower mass seeds, it is natural to expect that a leftover or progenitor population of IMBHs should also exist in the present universe, with various observable consequences. Several approaches to this question have been pursued.

### 5.1 Ultraluminous X-Ray Sources

Ultra-luminous (i.e., relative to "standard" X-ray binaries) X-ray sources (ULX) discovered by the *Chandra* mission (Fabbiano 1989) have been proposed as candidates for accreting IMBHs, simply on account of Eddington luminosity arguments (Colbert & Mushotzky 1999). Their status remains unclear; for reviews and references, see, e.g., Mushotzky (2004) or Makashima (2007), and other reviews in this volume.

### 5.2 Evidence from Stellar Dynamics in Globular Clusters

Suggestions that IMBHs may lurk in the cores of a few galactic globular clusters (GCs) date from the 1970's, when X-ray emission was first detected in GCs. However, as soon as the angular resolution improved, these X-ray sources were proven as being outside the cluster cores, and were soon identified as low-mass X-ray binaries and cataclysmic variables (CVs). A modern overview of this topic is given by Grindlay elsewhere in this volume.

Several recent claims for IMBHs are based on dynamical observations. Gerssen et al. (2002) mentioned an IMBH in the prototypical core collapse globular cluster M15, with $M_{BH} = (4.5 \pm 2.1) \times 10^3$ $M_{\odot}$. However, after the discovery of an error in another paper they had based part of their study on, Gerssen et al. (2003) scaled down their mass estimate to $M_{BH} = 1.7$ (+2.7 -1.7) $\times 10^3$ $M_{\odot}$ with the remark that models without a BH are now found to be statistically acceptable (within 1-σ), although inclusion of a BH still provides a marginally better fit. In their study of the same cluster, Baumgardt et al. (2003a) were able to fit the same data without an IMBH.

At about the same time, Gebhardt et al. (2002) detected an IMBH with $M_{BH} = (2.0 \pm 1.1) \times 10^4$ $M_{\odot}$ in the core of G1, the brightest GC in M31. In a reanalysis of these data, Baumgardt et al. (2003b) are again able to fit them without the need to invoke the presence of any IMBH. With new Keck and HST data on G1, Gebhardt et al. (2005) find an IMBH with $M_{BH} = (1.7 \pm 0.3) \times 10^4$ $M_{\odot}$. Unfortunately, none of these studies of G1 presents the essential (squared velocity dispersion vs. radius) diagram.

The best observational strategy to look for an IMBH in a GC is the choice of a nearby GC in which it is possible to measure high-quality kinematical data – proper motions and radial velocities – for numerous individual stars. McLaughlin et al. (2006) have acquired and analyzed such data for 47 Tucanae, a massive and highly concentrated cluster. This is the largest existing set of 13,000 high-quality proper motions, from 9-epoch HST observations scattered over a time baseline of 7 years. They also used 5,600 radial velocities obtained from ground-based facilities, although only a small fraction of them were useful. Among other interesting results, this study provides, for the first time, a direct kinematical evidence of mass segregation between two subpopulations : the heavier blue stragglers and the lighter red and subgiants.

Following the study by Tremaine et al. (2002), whose prediction corresponds, in the case of 47 Tucanae, to a sphere of influence $GM_{BH}/\sigma(0)^2 \approx 0.032$ pc = 1.6 arcsec (D/4kpc), a potential observational signature should be detectable in the McLaughlin et al. (2006) proper-motion sample.

By fitting simple models of isotropic, single-mass stellar clusters with central point masses to their observed $\sigma_\mu(R)$ profile, McLaughlin et al. (2006) infer a 1-$\sigma$ upper limit of $M_{BH} \leq 1000-1500$ $M_\odot$ for any intermediate-mass black hole in 47 Tucanae ; see Fig. 9. The formal best-fit hole mass ranges from 0 if only the kinematics of stars near the main-sequence turn-off mass are modeled, to $\sim 700-800$ $M_\odot$ if fainter, less massive stars are also used. They can neither confirm nor refute the hypothesis that 47 Tucanae might lie on an extension of the $(M_{BH}-\sigma)$ relation observed for galaxy bulges.

It will be challenging indeed to build more comprehensive velocity data sets, in 47 Tucanae or any other globular cluster, which might be able to overcome inherent statistical limitations and discern unambiguously the subtle kinematical effects of any central mass concentration with $M_{BH} \sim 10^3$ $M_\odot$ or less.

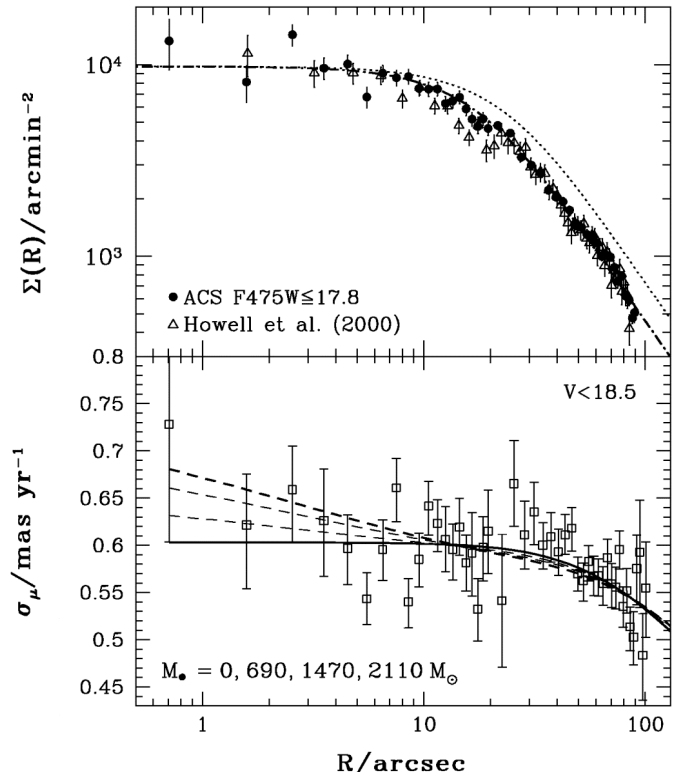

**Figure 9.** From a study of 47 Tucanae by McLaughlin et al. (2006). Top : Number density profile of stars brighter than the turnoff (dots) and compared with similar results from other sources (triangles). Bottom : proper-motion velocity dispersion profile. Curves represent fits of $W_0 = 8.6$ King models with central black holes of the mass indicated, which correspond to the formal best fit (solid curves) and the 68%, 95%, and 99% (uppermost dashed curve) upper limits.

Currently, there are two families of astrophysicists separated by the way they use the Ockham's razor : with the *same* observational results showing a marginally significant evidence, some of them conclude that the presence of IMBHs in Gcs cannot be excluded, while the others conclude that there is so far no conclusive evidence for IMBHs in GCs...

### 5.3  Low-Luminosity AGN, and Other Possibilities

Probably the most direct and straightforward evidence for IMBHs is in the nuclei of low luminosity AGN at low redshifts (Greene & Ho 2007ab, Ho 2008, and references therein). One complication in the interpretation of AGN data is that many fairly massive BHs are presently inactive, or can be accreting in a relatively inconspicuous manner, so a low luminosity AGN does not imply a lower mass BH. Nevertheless, low-luminosity AGN are natural homes for IMBHs.

Modulo the uncertainties of the nature of ULX sources, and the unpersuasive evidence from the dynamics of GCs, an implicit conclusion from all of these arguments is that there aren't very many "free floating" IMBHs left in the present-day universe, outside the galactic nuclei themselves. Granted, a "naked" IMBH floating in a galactic halo or the space between galaxies, with limited accretion possibilities, would be very difficult to detect; for a related discussion, see, e,g., Islam et al. (2004abc). Perhaps some limits to a large population of such objects can be derived from the statistics of gravitational (mini- or micro-) lensing, but they probably would not exclude modest numbers of such objects; for a related work, see, e.g., Wilkinson et al. (2001).

## 6    Concluding Comments and Future Prospects

Formation and early evolution of galaxies and black holes which grow to power quasars, their synergistic co-evolution and interplay with the intergalactic medium during the reionization era and after, are clearly among the most exciting subjects in cosmology today, and they touch upon numerous fundamental questions, as we hopefully started to convey in this review. There are also prospects for surprises and even detection of new astrophysical phenomena associated with the birth and growth of massive black holes.

The subject is growing rapidly, providing both observational and theoretical challenges, and will undoubtedly be a fruitful playground for the forthcoming generation of large telescopes on the ground and in space, and over the full range of wavelengths, and especially the nascent field of gravitational wave astronomy. Young black holes at large redshifts and their youthful shenanigans will certainly keep us busy for many years to come.

### Acknowledgments

We wish to acknowledge numerous collaborators, without whom we would not have had nearly as much fun (or results) in this field. Our apologies to any authors whose work we inadvertently neglected to cite – the field is vast, and our review spacetime was limited. We also wish to thank the conference organizers for their great efforts, and the saintly patience while waiting for this contribution. SGD is grateful to the staff of Palomar and Keck observatories for their expert help during numerous observing runs, and would like to acknowledge the stimulating atmosphere of the Aspen Center for Physics, where some of this work was done. We also acknowledge partial support from the U.S. NSF grant AST-0407448 and the Ajax Foundation (SGD), various grants (MV), Max-Planck Gesellschaft (VS), U.S. NSF grant AST-0708795 (VB), and Swiss National Science Foundation (GM).